\documentclass[12pt]{article}

\usepackage{pgfplots}
\usepackage{amssymb}
\usepackage[cmex10]{amsmath}
\usepackage{natbib}
\usepackage{multirow}

 \usepackage{enumerate}
\usepackage{fullpage}
\usepackage{pdfsync}
\usepackage{colordvi}
\allowdisplaybreaks
\usepackage[nodisplayskipstretch]{setspace}
\usepackage{microtype}
\usepackage{float}
\usepackage[caption = false]{subfig}
 \usepackage{booktabs}

\usepackage[T1]{fontenc}

\usepackage{lmodern}

\DeclareMathOperator*{\mmax}{\textsf{max}}
\DeclareMathOperator*{\ssup}{\textsf{sup}}
\newcommand{\E}{\mathbb{E}}

\newcommand{\Var}{\mathbb{V}}
\newcommand{\PP}{\mathbb{P}}

\newcommand{\Cov}{\mathbb{C}}

\newtheorem{proposition}{Proposition}
\newtheorem{corollary}{Corollary}
\newtheorem{lemma}{Lemma}

\numberwithin{equation}{section}

\usepackage{mathtools}
\usepackage{enumitem}
\usepackage{bm}

\newcommand{\hFe}{\hat{F}_e(e_i)}
\newcommand{\hhFe}{\hat{F}_{\hat{e}}(\hat{e}_i)}
\newcommand{\eps}{\varepsilon}
\newcommand{\tn}{\tilde n}

\usepackage{verbatim,color}
\definecolor{thistle}{rgb}{0.8,0.05,1}

\def\boxit#1{\vbox{\hrule\hbox{\vrule\kern6pt
          \vbox{\kern6pt#1\kern6pt}\kern6pt\vrule}\hrule}}

\title{Asymptotic Properties of Endogeneity Corrections Using Nonlinear Transformations}

\author{Jörg Breitung$^1$, Alexander Mayer$^2$ and Dominik Wied$^3$\mythanks{This Version: \today. $^1$,$^3$: Institute for Econometrics and Statistics, University of Cologne, e-mails: {\ttfamily breitung@statistik.uni-koeln.de}, {\ttfamily dwied@uni-koeln.de.} $^2$: Department of Economics, Universit\`{a} Ca' Foscari Venezia, e-mail: {\ttfamily alexandersimon.mayer@unive.it}.} \\[1mm]
}

\date{\today}

\usepackage{titlesec}
\titleformat{\section}[block]{\centering\normalfont\sffamily}{\thesection.}{0.5em}{\lsstyle\uppercase}
\titleformat{\subsection}[block]{\normalfont\sffamily}{\thesubsection.}{0.4em plus .1em minus .2em}{}
\titleformat{\subsubsection}[runin]{\normalfont\sffamily}{\thesubsubsection.}{0.4em plus .1em minus .2em}{}[.]
\titlespacing*\section{0pt}{18pt plus 4pt minus 2pt}{4pt plus 1pt minus 1pt}
\titlespacing*\subsection{0pt}{16pt plus 3pt minus 2pt}{4pt plus 1pt minus 1pt}
\titlespacing*\subsubsection{0pt}{12pt plus 2pt minus 1pt}{4pt plus 1pt minus 1pt}

\makeatletter
\def\mythanks#1{%
    \protected@xdef \@thanks {\@thanks \protect \footnotetext [\the \c@footnote ]{#1}}%
}
\makeatother

\usepackage{hyperref}
\begin{document}
\newtheorem{theorem}{Theorem}
\newtheorem{definition}{Definition}
\newtheorem{assumption}{Assumption}
\newtheorem{example}{Example}
\newtheorem{remark}{Remark}

\bibliographystyle{ecta}

{
\maketitle

   \begin{abstract}
This paper studies the asymptotic properties of endogeneity corrections based on nonlinear transformations without external instruments, which were originally proposed by Park and Gupta (2012) and have become popular in applied research. In contrast to the original copula-based estimator, our approach is based on a nonparametric control function and does not require a conformably specified copula. Moreover, we allow for exogenous regressors, which may be (linearly) correlated with the endogenous regressor(s). We establish consistency, asymptotic normality and validity of the bootstrap for the unknown model parameters. An empirical application on wage data of the US current population survey demonstrates the usefulness of the method.
\end{abstract}


\noindent \textbf{Keywords:} IV regression, generated regressors, wage equations
}

\section{Introduction}

Employing instrumental variables is a classical econometric tool for identifying causal effects if no randomized controlled trial or suitable observed control variables are available (see e.g. \citealp{angristpischke:2008}). Instruments allow for constructing quasi-experiments and yield consistent parameter estimators for endogenous regressors as long as they are uncorrelated with the error term. In empirical practice, however, a common problem is the choice of suitable instruments. As reviewed in \citet{ebbes:2009}, there are different approaches to construct consistent estimators in a regression with endogenous regressors without using (external) instruments, the `higher moments' approach (see \citealp{lewbel:1997}), the `identification through heteroscedasticity' approach (see \citealp{rigobon:2003}) and the `latent instrumental variable' approach (see \citealp{ebbes:2005}). Alternative approaches to instrument-free identification have been recently proposed by \cite{lewetal:22} or \cite{kiviet:23}.

The present paper follows up on \citet{guptapark:2012} who obtain identification via nonlinearity. The copula based endogeneity correction suggested by \citet{guptapark:2012} has recently become very popular in marketing research (see e.g. \citealp{burmeister:15}, \citealp{datta:2015}, \citealp{vomberg:20}, or \citealp{butt:21}) but also in economics (see e.g.  \citealp{aloui:16}, \citealp{blauw:16}, or \citealp{bhatt:21}). In contrast to the copula-based approach in earlier work, our endogeneity correction relies on a nonparametric control function approach (see e.g. \citealp{navarro:2010} and \citealp{wooldridge:2015} for an overview on control functions). We just assume that the endogeneity is due to an error component that enters the endogenous regressor in a nonlinear way. This approach does not need to assume that the nonlinear relationship between the endogenous regressor and the error term is known. Rather we just assume that it is monotonously increasing or decreasing. Hence our endogeneity adjustment is basically nonparametric, although the underlying regression model is fully parametric.\newline\hspace*{.35cm}
The idea goes back at least to \cite{guptapark:2012} who propose a related copula-based approach. In their estimation method, the joint distribution of the endogenous regressor and the error is represented by a copula. The parameters are estimated by maximum likelihood, where the likelihood function is based on the estimated joint distribution. There are two potential drawbacks in their approach. On the one hand, they do not study the asymptotic properties of their estimator and hence it is not clear under which conditions their approach is consistent and how to conduct statistical inference. On the other hand, they (implicitly) assume that any additional exogenous regressor is independent of the endogenous regressor. These drawbacks are partly addressed by \citet{haschka:2021} and \citet{yang:2022} who allow for a particular form of dependence between the exogenous and endogenous regressors. However, the proof for consistency in \citet{yang:2022} assumes that the distribution of the endogenous regressor is known which is typically not the case in empirical practice. Furthermore they do not study other asymptotic properties such as the convergence rate and asymptotic normality of their estimator. \citet{haschka:2022} provides an extension of the idea to linear panel models based on maximum likelihood, but does not study the asymptotic properties.\newline
\hspace*{.35cm}We study the asymptotic properties of a popular (control function based) variant of the estimator which does not require to specify a suitable copula. Furthermore, it allows for a correlation between the endogenous regressor and other exogenous regressors, and we rigorously study the asymptotic properties of our estimator. The estimator is calculated in two steps: First, we remove the (linear) dependence among the exogenous and endogenous regressors by computing the residuals from an auxiliary regression. Second, the endogenous part of the error is estimated by applying the quantile function of the normal distribution on the ranks of the residuals from the first step. In contrast, \citet{yang:2022} (2sCOPE) calculate residuals from a regression of the transformed regressors, thereby assuming a particular nonlinear dependence among the exogenous and endogenous regressors.\newline
\hspace*{.35cm}Due to the two-step estimation, it is not straightforward to derive asymptotic results for our estimator. A particular problem is that both the residuals and the ranks are dependent so that no simple law of large number or central limit theorem can be applied. We rely on results for stochastic integrals in order to analyze the asymptotic properties. In particular, we derive consistency and asymptotic normality of the structural parameters based on recent results on (residual) empirical process theory. The calculations reveal that the asymptotic distribution of the estimator depends on nuisance parameters arising from estimating a latent error term. This is similar in nature to the so-called `{\it generated regressor issue}' originating from the work of \cite{pagan:1984} and which has since then been discussed elsewhere in the literature, see, among others, \cite{chen2003estimation}, \cite{hahnridder:13}, \cite{mammenetal:12}, or \cite{mammenetal:16}.\newline
\hspace*{.35cm}The plan of our paper is as follows. In Section \ref{sec:2}, we develop our empirical endogeneity correction, discuss connections to former copula-based approaches as well as to instrumental variable estimation. Asymptotic properties are derived in Section \ref{sec:asyProp}. Next, we study the small sample properties by means of Monte Carlo simulation in Section \ref{sec:Simulations} and provide an empirical application to wage data (Section \ref{sec:Application}). This application is well suited for our approach because the potentially endogenous regressor (years of education) is apparently not normal and because there is an ongoing debate in the empirical literature about the choice of appropriate instruments.

\section{Model, Identification, and Estimator}\label{sec:2}

\noindent Consider a structural linear model
\begin{align}\label{model}
y = \beta'x + \gamma z + u,
\end{align}
where \(x\) is a \(k \times 1\) vector of exogenous regressors (including a constant) and \(z\) is a scalar endogenous regressor correlated with the error term \(u\). Let us further assume that the endogenous variate and the error can be decomposed as
\begin{align}\label{udecomp}
z = \delta'x + e \quad \text{ and } \quad u = \rho f(e) + \eps,
\end{align}
where \(e\) and \(\eps\) are mean-zero error terms satisfying $\E[\eps\mid x,z] = 0$ and $e \perp x$. If the function $f$ is linear, the parameter $\rho$ is not identified in general without using external instrumental variables. \citet{lewetal:22} consider a somewhat different triangular system in which $\rho$ is identified without external instrumental variables if the latent error terms are not normally distributed and the sign of $\rho$ is fixed. We pursue a different path, i.e. we can allow for normally distributed \(f(e)\) and \(\eps\), but require that the function \(f(\cdot)\) is nonlinear so that \(e\) must not be Gaussian. If the function $f(\cdot)$ is nonlinear and known, then it is possible to identify and correct for endogeneity by just augmenting the regression by $f(e)$ and in this case we refer to $f(e)$  as the `control function'. This result is summarized by the following theorem.

\begin{theorem}\label{thm:ident}
    Assume that Eqs.\ \eqref{model} and \eqref{udecomp} hold, with $\E[\eps\mid x,z] = 0$, $e \perp x$, $\Var[e]>0$, $f(\cdot)$ striclty monotone, and $f(e) \sim \mathcal{N}(0, 1)$.\footnote{Since we can absorb mean and variance of \(f(e)\) into the intercept and \(\rho\) in Eqs. \eqref{model} and \eqref{udecomp}, we shall, without loss of generality, assume in the following that \(\E[f(e)] = 0\) and \(\Var[f(e)] = 1.\)} Then, $(\beta',\gamma,\rho)'$ are identified if and only if the distribution of $e$ is not normal.
\end{theorem}

\noindent The setting is mainly inspired by \cite{guptapark:2012} who consider the special case of independence between endogenous and exogenous regressor, i.e. \(\delta = 0\) in Eq. \eqref{udecomp}. Under the assumption that the marginal cumulative distribution function (c.d.f.) \(F_u(t) \coloneqq \PP(u \leq t)\) as well as the copula governing the joint distribution of \((u,e)\) are normal, they propose an estimation method using a Gaussian copula. Since the resulting likelihood depends on the unknown c.d.f. \(F_e(t) \coloneqq \PP(e \leq t)\), the latter is estimated by integrating a kernel density estimator. To illustrate, assume for simplicity that \(\Var[e] = \Var[u] = 1\) and let \(\{y_i,x_i,e_i\}_{i = 1}^n\) be a sample of length \(n\) of independent and identically distributed (\textit{i.i.d.}) copies of \((y,x,e)\). The resulting estimator of the true parameter \(\alpha_0 \coloneqq (\beta_0',\gamma_0)'\), say, maximizes the approximate log-likelihood
\begin{align*}
\ell_n(\alpha; \hat{F}_e&) \\
\coloneqq \,&
\frac{n}{2}\textsf{ln} (1-\rho^2)  + \sum_{i=1}^n\textsf{ln}\phi(u_i(\alpha))\\
\,&  +  \sum_{i=1}^n\frac{\rho^2 \left[ \Phi^{-1}(\hat{F}_e(e_i))^2 + \Phi^{-1}(F_u(u_i(\alpha)))^2 \right] - 2\rho  \Phi^{-1}(\hat{F}_e(e_i)) \Phi^{-1}(F_u(u_i(\alpha)))}{2(1-\rho^2)},
\end{align*}
where \(u_i(\alpha) \coloneqq y_i - \beta'x_i - \gamma z_i\) is a generalized residual, \(\Phi(\cdot)\) and \(\phi(\cdot)\) are the standard normal c.d.f. and density, respectively, \(H^{-1}\) is the left-continuous generalized inverse of any c.d.f. \(H\),  and we use the notation \(\ell_n(\alpha; \hat{F}_e)\) to indicate that the log-likelihood depends on the nonparametric estimator \(\hat{F}_e\) of the unknown distribution \(F_e\). Since the (approximate) likelihood function is quite complicated, it needs to be maximized by numerical techniques.

Notwithstanding its great conceptual value, this procedure also has at least two major drawbacks. First, the approach does not account for dependence between endogenous and exogenous regressors. Second, and perhaps the most important practical concern, it is \textit{a priori} unclear under which assumptions the usual properties of ML estimation carry over (if at all) to the case with nonparameterically estimated infinite dimensional nuisance parameter \(F_e\); see e.g. \cite{genest:1995}. As we demonstrate below, deriving precise statements about limiting properties in the presence of a nonparameterically generated regressor is a highly non-trivial undertaking. Thus, there is no asymptotic theory in \cite{guptapark:2012} to enable and/or justify statistical inference.

In order to overcome these issues, we propose an easy-to-implement alternative based on a simple augmented regression that does not require the specification of the joint distribution \((u,e)\). Akin to \citet{guptapark:2012}, we assume that there exists a strictly monotonic and nonlinear function \(f(\cdot)\) such that $f(e)$ is distributed according to some c.d.f. $F_e$. In the following, we assume that $f(e)$ is Gaussian, which is a plausible assumption in many applications. For example, in our empirical application below, $f(e)$ represents latent information about personal information such as intelligence, other skills or diligence, compare the discussion in Section \ref{sec:Application}. It then follows by standard arguments that \(\Phi^{-1}(F_e(e)) = f(e)\). Hence, if \(e\) and its distribution \(F_e\) were known, then one can identify \(\beta\) and \(\gamma\) by augmenting Eq. \eqref{model} with \(\eta \coloneqq \Phi^{-1}(F_e(e))\); see Theorem \ref{thm:ident}. 


 Note that these assumptions are similar but not identical to the assumptions of  \citet{guptapark:2012}. For example, whereas in Eq. \eqref{udecomp} we assume that $f(e)$ is normal (or has some other known distribution),  \citet{guptapark:2012} assume that $u$ is normally distributed. For ensuring the latter, one needs to assume that also $\eps$ is normally distributed, which we do not. Another difference is that we allow for \(\delta \neq 0\), while this sort of dependence between \(z\) and \(x\) has been ruled out by \citet{guptapark:2012}. Moreover, and in contrast to the aforementioned study, we allow for conditionally heteroscedasticity of $\eps$.

In practice, \(F_e\) is rarely known and has to be estimated. Here, we approach this problem nonparametrically using the empirical distribution. That is, we use the ordinary least-squares (OLS) estimator for Eq. \eqref{model} augmented with a feasible sample counterpart of \(\eta\). More specifically, in a preliminary step the (linear) dependence between \(z\) and \(x\) is eliminated using the OLS estimator \(\hat\delta\) in the regression of \(z\) on \(x\). From the first-stage residuals \(\hat e_i \coloneqq z_i - \hat\delta'x_i\), \(i \in \{1,\dots,n\},\) we then compute the empirical distribution and obtain the following nonparametrically generated regressor
\begin{align}
\widehat \eta_i \coloneqq \Phi^{-1}(\hhFe),\label{ranktransformation}
\end{align}
where $\hhFe$ denotes the (rescaled) empirical distribution function of $e$, i.e.
\begin{align*}
\hhFe \coloneqq \frac{1}{n+1}\sum_{j=1}^n1\{\hat e_j \leq \hat e_i\} = \frac{\textsf{rank of } (\hat e_i)}{n+1}.
\end{align*}
We adopt common practice and rescale by \(n+1\) in order to escape the possibility of $\hhFe=1$ (note that $\Phi^{-1}(a)\to \infty$ as $a\to 1$).\footnote{Note that dividing by $n+1$ instead of $n$ is innocuous as  \[\ssup\limits_{t \in \mathbb{R}}|\hat{F}_{\hat e}(t)-\tilde{F}_{\hat e}(t)| \leq 1/(n+1),\] where \(\tilde{F}_{\hat e}(\cdot) \coloneqq (n+1)\hat{F}_{\hat e}(\cdot)/n \).
Moreover, note that the fact that $\Phi^{-1}(a)\to -\infty$ as $a\to 0$ does not yield any difficulties in the definition of the estimator as $n \geq 1$.} Finally, we can estimate \(\theta \coloneqq (\beta',\gamma,\rho)'\) using the OLS estimator of the model in Eq. \eqref{model} augmented by \(\widehat \eta_i\).

\begin{remark}
Assuming that the endogeneity enters the error term in an additive manner, it is relatively easy to extend our approach in order to accommodate further endogenous regressors. To fix ideas, consider the special case of two endogenous regressors $z_1$ and $z_2$ so that
\[
y = \beta'x + \gamma_1z_1+\gamma_2z_2+u, \quad u = \rho_1f_1(e_1)+\rho_2f_2(e_2) + \varepsilon,
\]
where $z_j = \delta_j'x+e_j$, $j \in \{1,2\}.$ Let $F_j(\cdot)$ be the c.d.f. of $e_j$, $j \in \{1,2\}$. Assuming that $f_j(e_j)$ are normally distributed and $f_j(\cdot)$ are strictly monotone, one obtains 
\[
f_j(e_j) = \Phi^{-1}(F_j(e_j)) \eqqcolon \eta_j,
\]
say.  For this additive model specification we can easily replace $f_j(e_j)$ by $\hat\eta_{j}$, where $\hat\eta_{j}$ is constructed from the emprical c.d.f. of residuals from individual first stage regressions of $z_j$  on $x$. Thus, we could estimate the $(k+4) \times 1$ parameter vector $$\theta \coloneqq (\beta',\gamma',\rho')', \quad \gamma \coloneqq (\gamma_1,\gamma_2)',\; \rho \coloneqq (\rho_1,\rho_2)'$$ from the augmented regression
$y = \beta'x + \gamma'z + \rho'\hat\eta + \textnormal{\sf error}$, $z \coloneqq (z_1,z_2)'$, $\hat\eta \coloneqq (\hat\eta_1,\hat\eta_2)'$. If the vector $(x,z,e,\eps)$ satisfies correspondingly \textnormal{Assumption \ref{ass:A}} specified below, then our subsequent results carry over.
\end{remark}

\begin{remark}
It is also possible to relate our approach to the standard instrumental variable estimation. To simplify the exposition we focus on the model with a single endogenous regressor which we write in vector format as
\(y = z\gamma + \widehat \eta \rho + \tilde \varepsilon,\) with \(\varepsilon \coloneqq \eps + \rho(\eta - \hat\eta),\)
where $y\coloneqq(y_1,\ldots,y_n)'$, $z\coloneqq(z_1,\ldots,z_n)'$, $\widehat \eta \coloneqq (\widehat \eta_1,\ldots,\widehat \eta_n)'$. The estimator can be represented as
\begin{align*}
\widehat \gamma & \coloneqq \frac{ z'M_{\hat \eta} y }{ z'M_{\hat \eta} z } ~=~ \frac{ \widehat v'y }{ \widehat v'z }
\end{align*}
where $\widehat v \coloneqq M_{\hat \eta} z$ and $M_{\hat \eta} \coloneqq I_n - \widehat \eta \widehat \eta'/\widehat \eta'\widehat \eta$. This representation of the estimator gives rise to the interpretation of a just-identified IV estimator using the residuals from a regression of $z$ on $\widehat \eta$ as instrumental variable vector $\widehat v$. Accordingly, for the full model we may alternatively estimate the coefficients $\beta$ and $\gamma$ from an IV regression using $(x_i, \widehat v_i)$ as instruments with $\hat v_i$ as the $i$'th element of the residual vector $\widehat v$. This allows us to combine the `internal instrument' $\widehat v_i$ with possible external instruments resulting in an over-identified IV \textnormal{(}or GMM\textnormal{)} estimator. Furthermore the approach of \textnormal{\citet{stockyogo:2005}} may be adapted to test against `weak instruments'. It should be noted, however, that $\widehat v_i$ is an estimated instrumental variable and its estimation error affects the asymptotic distribution, as will be shown in more detail below using recent findings on residual empirical processes.
\end{remark}

\section{Asymptotic Properties} \label{sec:asyProp}

Consider the joint OLS estimator of \(\theta \coloneqq (\beta',\gamma,\rho)'\) given by \(\hat\theta \coloneqq (W'W)^{-1}W'y\), where \(y \coloneqq (y_1,\dots,y_n)'\) and \(W \coloneqq (w_1',\dots,w_n')'\) is a \((k+2)\times n\) matrix for \(w_i \coloneqq (x_i',z_i,\hat{\eta}_i)'\), \(i \in \{1,\dots,n\}\). We aim at deriving the asymptotic distribution of
\begin{align}\label{eq:thetadecom}
\sqrt{n}(\hat\theta-\theta) = (W'W/n)^{-1}\frac{1}{\sqrt{n}}(W'\eps+\rho W'(\eta-\hat\eta)),
\end{align}
where \(\eta \coloneqq (\eta_1,\ldots,\eta_n)'\) and \(\hat\eta \coloneqq (\hat\eta_1,\ldots,\hat\eta_n)'\). Technically, this represents a nontrivial task because of the term \(W'(\eta-\hat\eta)/\sqrt{n}\) that arises due to a nonparametrically generated regressor. The  presence of these normal quantile transforms of estimated residuals considerably complicates the derivation of an appropriate asymptotic theory. We approach this problem by combining seminal results on empirical quantile process theory (see e.g. \citealp{koenkerxiao:2002}) and stochastic calculus with recent findings of \citet{zhao:2020} on residual rank approximations. In order to do so, we have to impose some assumptions.

\renewcommand{\theassumption}{A}
\begin{assumption}\label{ass:A}\textcolor[rgb]{1,1,1}{.}
\begin{enumerate}[label=\textnormal{A\arabic*},ref=A\arabic*]
\item\label{ass:A0} \(\{y_i,x_i,z_i\}_{i=1}^n\) is a sample of i.i.d. copies of \((y,x,z)\).
\item\label{ass:Aa} $\sum_{i=1}^nx_ix_i'/n \stackrel{p}{\rightarrow} \E[xx'] \eqqcolon \Sigma_x$, and  $\sum_{i=1}^nx_ix_i'\eps_i^2/n \stackrel{p}{\rightarrow} \E[xx'\eps^2]$ where $\Sigma_x$ and $\E[xx'\eps^2]$ are \(k \times k\) positive definite matrices and  $\E[x] \eqqcolon\mu_x$.
\item\label{ass:Ab} There exists a strictly monotonous function $f(\cdot)$ such that $f(e) \sim {\cal N}(0,1)$.
\item\label{ass:Ac} There exists a decomposition of the endogenous regressor such that $z = \delta'x + e$,  where $e$ is independent of $x$, with \(\E[e] = 0\), \(\Var[e] \eqqcolon \sigma_e^2 > 0\), and $\E[e^4]<\infty$.
\item\label{ass:Ad} \textnormal{(\textit{i})} \(e\) has differentiable c.d.f. \(F_e\) that does not coincide with the normal distribution. \textnormal{(\textit{ii})} For the density $f_e$ of $e$, the following must hold: There is a constant $\gamma \in (1/2,1)$ such that, for $a \rightarrow 0$
\begin{equation*}
\sup_{u \in (a,1-a)} \frac{f_e(F^{-1}_e(u))}{\mathsf{min}(u,1-u)} = o\left(a^{-1/(2\gamma)} \right).
\end{equation*}
\item\label{ass:Ae} $\E[\varepsilon \mid x,z]=0$, $\E[\eps^2 \mid x,z] \in (0,\infty)$ a.s., and $\E[\eps^4] <\infty$.
\end{enumerate}
\end{assumption}

\noindent The first part of Assumption \ref{ass:Ab} is similar to the first-stage monotonicity condition employed frequently in the literature on causal inference (see e.g. \citealp{imbang:1994} or \citealp{imbnew:2009}) and justifies the representation \(\Phi^{-1}(F_e(e)) = f(e)\). Assumption \ref{ass:Ac} implies that $x$ affects $z$ only linearly. In applications, this limitation can be addressed by adding nonlinear transformations of the exogenous regressors to $x$. We could, in principle, relax this assumption and model the relation between exogeneous and endogenous regressor nonparametrically using recent results of \citet{zhao:2022}. Since deriving an asymptotic theory is already fairly complex, we leave this issue to future research. The first part of Assumption \ref{ass:Ad} is essential for identification as mentioned above. Without any further specification of the distribution of \(\eps\)  the normality of \(f(e)\) is not directly testable; see also the discussion in \cite{becker:22}. Still, in practice one could use the first-stage residuals \(\hat e\) to test whether \(e\) is Gaussian, which would lead to non-identification. The second part of Assumption \ref{ass:Ad} requires that the density of $e$ decays sufficiently fast to $0$ at the boundary. The condition is satisfied by many parametric distributions like the Gamma distribution with shape parameter of at least two or the Chi-square distribution with degrees of freedom parameter of at least three. This is a technical requirement needed to deal with the estimation error of the normal scores introduced by the first-stage residuals; see \citet[Assumption 3.5]{zhao:2020} for details. Finally, the mean-independence in Assumption \ref{ass:Ae} ensures that $\eps$ is orthogonal to all regressors in the augmented regression, while conditional heteroscedasticity is allowed for.

\begin{proposition}\label{prop:PA} If Assumption \ref{ass:A} is satisfied, then
\(
\sqrt{n}(\hat\theta-\theta) \stackrel{d}{\rightarrow} \mathcal{N}(0_{k+2},\Sigma), \quad \Sigma \coloneqq M^{-1}\Omega M^{-1},
\)
where
\[
M \coloneqq  \begin{bmatrix} \Sigma_x & \Sigma_x'\delta & 0_{k} \\ \delta'\Sigma_x & \delta'\Sigma_x\delta+\sigma_e^2 & c_2 \\ 0_{k}' & c_2 & 1 \end{bmatrix}\quad \text{ and } \quad
\Omega \coloneqq  \begin{bmatrix} \Omega_1 				&  \Omega_1\delta																			 & 0_{k} \\
																  \delta'\Omega_1 & \delta'\Omega_1\delta+\omega_1 & \omega_{12} \\
																	0_{k}' 																				&\omega_{12}																												 & \omega_2\end{bmatrix},
\]
with \(\Omega_1 \coloneqq \E[xx'\eps^2]+\Sigma_x\rho^2c_1^2\sigma_e^2\), \(\omega_1 \coloneqq  \E[e^2\eps^2]+\rho^2c_3\), \(\omega_2 \coloneqq \E[\eta^2\eps^2]+\rho^2/2\), $\omega_{12} \coloneqq \E[e\eta\eps^2] + \rho^2c_2/2$, and
\[
c_1 \coloneqq \int_0^1\frac{f_e(F_e^{-1}(u))}{\phi(\Phi^{-1}(u))} du, \quad  c_2 \coloneqq \int_0^1F_e^{-1}(u)\Phi^{-1}(u) du,
\]
and
\[\normalfont
c_3 \coloneqq \int_0^1\int_0^1  \frac{F_e^{-1}(u)}{\phi(\Phi^{-1}(u))}\frac{F_e^{-1}(v)}{\phi(\Phi^{-1}(v))} (\textsf{min}(s,u)-su) ds du.
\]
\end{proposition}

\noindent Proposition 3,1 reveals that the limiting distribution is affected by finite (\(\delta\)) and infinite dimensional (\(F_e(\cdot)\)) nuisance parameters that arise due to not knowing the (linear) relationship between exogenous and endogenous regressors and the unknown functional form of the endogeneity, respectively. One way in which this so-called `generated regressor' issue (see e.g. \citealp{pagan:1984}) could be mitigated consists in expressing our two-step estimator in terms of a (joint) moment estimator whose moment conditions are orthogonalized using suitable influence functions as proposed recently by \cite{cheetal:23}. Finally, we note that the identification failure of \(e\) being Gaussian leads to \(c_j = 1\), \(j \in \{1,2,3\}\), so that \(M\) is not invertible.

The limiting variance-covariance matrix could be estimated by a sample analogy principle. The only parameters difficult to estimate are \(c_j\),  \(j \in \{1,2,3\}\); but they could, in general, be replaced by sample counterparts using \(\hat{F}_{\hat{e}}(\cdot)\) and a suitable density estimator
\[
\hat{f}_{\hat{e}}(y) \coloneqq \frac{1}{nh_n}\sum_{i = 1}^nK\left(\frac{\hat{e}_i-y}{h_n}\right),
\]
say, where \(K(\cdot)\) is a Kernel density function and \(h_n\) is the bandwidth such that \(h_n \rightarrow 0\)  as \(n \rightarrow \infty\), see e.g. \cite{hardlint:1994}. Given all the difficulties involved in density estimation and its oftentimes poor performance in finite samples, we refrain from this idea and resort to bootstrap standard errors instead. We propose a resampling scheme that is able to account for the sampling uncertainty of the nonparametrically generated regressor thereby reproducing the correct distribution of Proposition 1. Similar to the control function literature (see e.g. \citealp{wooldridge:2015}), we simply generate bootstrap samples by drawing with replacement from the original data; see also \cite{chenetal:03} for a more general treatment of bootstrap inference with nonparametrically estimated nuisance parameters. More specifically, let \({\cal X}^*_1,\dots,{\cal X}^*_n\) be drawn with replacement from the empirical distribution of the vectors $\{ {\cal X}_1,\ldots,{\cal X}_n\},$ ${\cal X}_i \coloneqq (y_i,x_i',z_i)'$, and define, analogously to \(\hat\theta\), \(\hat\theta^*\) based on the bootstrap data. Corollary 1 below then justifies the use of bootstrap (percentile) confidence intervals as well as standard errors constructed from a large number of the bootstrap estimators \(\hat\theta^*\).
\begin{corollary}\label{cor:CA0} \textnormal{(\textit{i})} If Assumption \ref{ass:A} is satisfied, then \(\sqrt{n}(\hat\theta^*-\hat\theta) \stackrel{d}{\rightarrow} \mathcal{N}(0_{k+2},\Sigma)\) under the probability measure \(\mathbb{P}_{\cal X}\) implied by the bootstrap. \textnormal{(\textit{ii})} If, in addition, there exists some \(\nu > 0\) such that \(\E^*[\lVert\sqrt{n}(\hat\theta^*-\hat\theta)\rVert^{2+\nu}] < \infty\), then \(n\E^*[(\hat\theta^*-\hat\theta)(\hat\theta^*-\hat\theta)'] \rightarrow_{\mathbb{P}_{\cal X}} \Sigma\),  where \(\E^*\) is the expectation induced by \(\mathbb{P}_{\cal X}\).
\end{corollary}

\noindent The moment condition in part ($ii$) of Corollary 3.1 could, in principle, be derived from more primitive assumptions. If this condition fails, test statistics equipped with bootstrap standard errors tend to bee too conservative, as shown recently by \cite{hahnliao:2021}.

Although, it is generally true that hypothesis tests equipped with naive OLS standard errors yield asymptotically inaccurate results, an exception is given for a Durbin-Hausman-Wu type-test of the null hypothesis \(H_0\): \(\rho = 0\) of no endogeneity. More specifically, as summarized by Corollary 3.2 below, the usual approach based on a textbook \(t\)-statistic remains valid; a situation reminiscent of the `\textit{generated regressors}' literature, see e.g. \cite{pagan:1984}.

\begin{corollary}\label{cor:CA} Let Assumptions \ref{ass:A} be fulfilled. Under $H_0:$  $\rho=0$ the regressor $z$ is exogenous and the respective $t$-statistic has a standard normal limiting distribution as $n\to \infty$.
\end{corollary}
 
\section{Small Sample Properties}\label{sec:Simulations}

\begin{figure}[htbp]
\begin{center}
\begin{tikzpicture} 
\begin{axis}[
	tick label style={font=\small},
	label style={font=\small},
	tick label style={font=\small},
	legend style={font=\small},
  ymin=0,
	ymax=1,
	xmin=0,
	xmax=8,
	 xtick={0,2,4,6,8}, 
	 ytick={0,.5,1}, 
 xtick align=inside,
 xtick pos=upper,xticklabel pos=upper,
  scale = 0.9,
]
\addplot[style=solid, thick,  color = black] table[x="x",y="y1"] {gamma.txt}; 
\addplot[style=solid, dashed, color = black] table[x="x",y="y3"] {gamma.txt};

\node[scale=0.72, align=left]  at (65, 90)  {$\Gamma(1,1)$};
\node[scale=0.72, align=left]  at (180, 52)  {$\Gamma(3,2)$};
\end{axis}
\end{tikzpicture}
\end{center}\vspace*{-.8cm}
\caption{\small $\Gamma$-densities}\label{fig:densities}
\vspace*{-.2cm}
\end{figure}
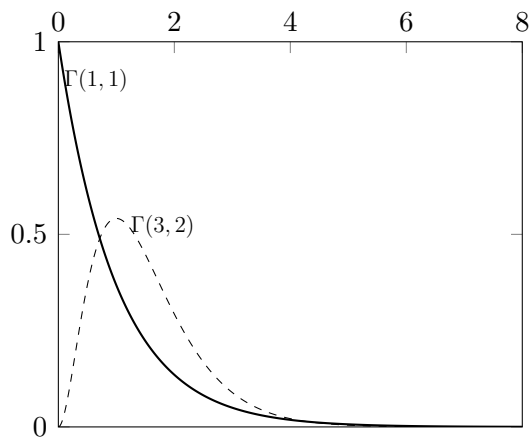

In order to investigate the small sample properties of our estimator, we consider a simple linear model inspired by \cite{yang:2022}:
\begin{equation}
\begin{split}
y = & \,\beta_0 + \beta_1 x + \gamma z + u,  \\
z = & \,\delta x + e, \qquad \text{where} \qquad  x, e \sim \Gamma(a,b),\; a,b > 0,
\end{split}
\end{equation}
 with $(\beta_0,\beta_1,\gamma)'=(1,-1,1)'.$ To this end, we follows \cite{yang:2022} by setting $x \sim \Gamma(1,1)$. Moreover, we distinguish between two different (gamma) distributions for the component $e$, i.e. ($i$) $\Gamma(1,1)$ or ($ii$) $\Gamma(3,2)$. Although ($i$) violates Assumption \ref{ass:Ad}, we report results to alleviate comparison with \cite{yang:2022}. As can be deduced from Figure \ref{fig:densities}, passing from ($ii$) to ($i$), the distribution of $e$ (and therefore $z$) becomes highly skewed. A more challenging situation results if $e \sim \Gamma(3,2)$ as this distribution is more similar to that of a Gaussian random variable. For the remaining distributional specifications, two different data generating processes are considered ({\sf DGP1} and {\sf DGP2}).

\noindent {\sf DGP1.} According to Eq. \eqref{udecomp}, the error term is generated as 
\(u = \rho \eta + \varepsilon,\) $\varepsilon \sim {\cal N}(0,1)$, and the standard normal distributed random variable $\eta$ is given by $\eta = \Phi^{-1}(F_{e}(e))$, where $F_{e}$ is the c.d.f. of the respective $\Gamma$-distribution specified above. We distinguish the cases of uncorrelated ($\delta=0$) and correlated ($\delta=1$) regressors as well as of no ($\rho=0$), moderate ($\rho=0.5$) and strong ($\rho=0.9$) endogeneity.\newline

\noindent {\sf DGP2.} The second design obtains with $\delta = 0$ and is taken from \citet[Section 4.1]{yang:2022} who assume that 
\[
e = F_{e}^{-1}(\Phi(e^*)),\; x = F_{x}^{-1}(\Phi(x^*)),\quad
(e^*,x^*,u)' \sim \mathcal{N}(0_3,\Xi), \;  \Xi \coloneqq \begin{bmatrix} 1 & \alpha & \rho \\ \alpha & 1 & 0 \\ \rho& 0 & 1 \end{bmatrix},
\]
where, as pointed out above, $F_{e}$ varies across the different gamma distributions and $F_x = \Gamma(1,1)$. The endogeneity and the dependence between $x$ and $e$ is captured by the correlation coefficients $\rho$ and $\alpha$, respectively. Again, we distinguish the cases of uncorrelated ($\alpha=0$) and correlated ($\alpha=1/2$) regressors as well as of no ($\rho=0$) or strong ($\rho=1/2$) endogeneity.

\begin{table}[htbp]
\footnotesize 
\begin{center}
\caption{\small \sf DGP1}\label{table1} 
\addtolength{\tabcolsep}{-0.6em}
\begin{tabular}{clcccccccccccccccc} \toprule
&& \multicolumn{8}{c}{$\Gamma(1,1)$} & \multicolumn{8}{c}{$\Gamma(3,2)$}\\
 \cmidrule(l){3-10} \cmidrule(l){11-18} 
&& \multicolumn{4}{c}{$\beta $} &  \multicolumn{4}{c}{$\gamma $} & \multicolumn{4}{c}{$\beta $} &  \multicolumn{4}{c}{$\gamma $}\\
 \cmidrule(l){3-6} \cmidrule(l){7-10} \cmidrule(l){11-14} \cmidrule(l){15-18}
\multicolumn{2}{l}{$\delta=0$} & \sf OLS  &   \sf GP   &  \sf npCF  & \sf 2sC &   \sf OLS  &  \sf GP   & \sf npCF  & \sf 2sC & \sf OLS  &   \sf GP   &  \sf npCF  & \sf 2sC &   \sf OLS  &  \sf GP   & \sf npCF  & \sf 2sC\\ \hline
\parbox[t]{2mm}{\multirow{4}{*}{\rotatebox[origin=c]{90}{$\rho =0.0$}}}&	\textsf{bias}&	0.003	&	0.003	&	0.003	&	0.003	&	-0.001	&	-0.004	&	-0.004	&	-0.004	&	0.003	&	0.003	&	0.003	&	0.003	&	-0.001	&	-0.008	&	-0.008	&	-0.008	\\
&	\textsf{std}	&	0.064	&	0.064	&	0.066	&	0.064	&	0.064	&	0.167	&	0.153	&	0.149	&	0.063	&	0.064	&	0.066	&	0.065	&	0.064	&	0.271	&	0.248	&	0.242	\\
&	\textsf{rmse}	&	0.064	&	0.064	&	0.066	&	0.064	&	0.064	&	0.167	&	0.153	&	0.149	&	0.064	&	0.064	&	0.066	&	0.065	&	0.064	&	0.271	&	0.248	&	0.243	\\
&	\textsf{size}	&	0.052	&	0.057	&	0.047	&	0.058	&	0.062	&	0.044	&	0.053	&	0.045	&	0.051	&	0.057	&	0.043	&	0.055	&	0.054	&	0.042	&	0.051	&	0.047	\\
\cmidrule(l){3-10}\cmidrule(l){11-18}	
\parbox[t]{2mm}{\multirow{4}{*}{\rotatebox[origin=c]{90}{$\rho =0.5$}}}&	\textsf{bias}&	0.004	&	0.004	&	0.003	&	0.003	&	0.451	&	-0.021	&	0.014	&	0.010	&	0.004	&	0.004	&	0.002	&	0.003	&	0.481	&	0.060	&	0.017	&	0.023	\\
&	\textsf{std}	&	0.065	&	0.064	&	0.086	&	0.070	&	0.066	&	0.175	&	0.16	&	0.155	&	0.064	&	0.064	&	0.076	&	0.071	&	0.067	&	0.282	&	0.259	&	0.252	\\
&	\textsf{rmse}	&	0.065	&	0.064	&	0.086	&	0.070	&	0.456	&	0.177	&	0.161	&	0.156	&	0.064	&	0.064	&	0.076	&	0.071	&	0.486	&	0.288	&	0.259	&	0.253	\\
&	\textsf{size}	&	0.051	&	0.058	&	0.047	&	0.053	&	1.000	&	0.047	&	0.055	&	0.048	&	0.050	&	0.059	&	0.035	&	0.045	&	1.000	&	0.057	&	0.051	&	0.044	\\
\cmidrule(l){3-10}\cmidrule(l){11-18}							
\parbox[t]{2mm}{\multirow{4}{*}{\rotatebox[origin=c]{90}{$\rho =0.9$}}}&	\textsf{bias}&	0.004	&	0.004	&	0.003	&	0.004	&	0.813	&	-0.034	&	0.029	&	0.022	&	0.004	&	0.004	&	0.002	&	0.003	&	0.867	&	0.113	&	0.038	&	0.047	\\
&	\textsf{std}	&	0.068	&	0.065	&	0.117	&	0.082	&	0.071	&	0.191	&	0.173	&	0.169	&	0.065	&	0.064	&	0.092	&	0.081	&	0.073	&	0.302	&	0.279	&	0.271	\\
&	\textsf{rmse}	&	0.068	&	0.065	&	0.118	&	0.082	&	0.816	&	0.194	&	0.175	&	0.17	&	0.065	&	0.064	&	0.092	&	0.081	&	0.87	&	0.323	&	0.281	&	0.275	\\
&	\textsf{size}	&	0.069	&	0.058	&	0.056	&	0.054	&	1.000	&	0.045	&	0.051	&	0.048	&	0.055	&	0.056	&	0.042	&	0.052	&	1.000	&	0.083	&	0.056	&	0.052	\\
\vspace*{-.32cm}\\
\cmidrule(l){3-10}\cmidrule(l){11-18}
\multicolumn{2}{l}{$\delta=1$} & \sf OLS  &   \sf GP   &  \sf npCF  & \sf 2sC &   \sf OLS  &  \sf GP   & \sf npCF  & \sf 2sC & \sf OLS  &   \sf GP   &  \sf npCF  & \sf 2sC &   \sf OLS  &  \sf GP   & \sf npCF  & \sf 2sC\\ \hline
\parbox[t]{2mm}{\multirow{4}{*}{\rotatebox[origin=c]{90}{$\rho =0.5$}}}&	\textsf{bias}&	-0.451	&	-0.453	&	-0.009	&	-0.235	&	0.640	&	0.333	&	0.019	&	0.305	&	-0.554	&	-0.543	&	-0.016	&	-0.409	&	0.735	&	0.489	&	0.025	&	0.523	\\
&	\textsf{std}	&	0.095	&	0.096	&	0.185	&	0.166	&	0.100	&	0.248	&	0.231	&	0.227	&	0.099	&	0.100	&	0.312	&	0.189	&	0.109	&	0.311	&	0.401	&	0.255	\\
&	\textsf{rmse}	&	0.461	&	0.463	&	0.185	&	0.288	&	0.648	&	0.415	&	0.231	&	0.38	&	0.562	&	0.552	&	0.313	&	0.451	&	0.743	&	0.579	&	0.402	&	0.582	\\
&	\textsf{size}	&	0.997	&	0.996	&	0.041	&	0.303	&	1.000	&	0.246	&	0.055	&	0.255	&	1.000	&	0.999	&	0.047	&	0.570	&	1.000	&	0.353	&	0.051	&	0.187	\\
\cmidrule(l){3-10}\cmidrule(l){11-18}
\parbox[t]{2mm}{\multirow{4}{*}{\rotatebox[origin=c]{90}{$\rho =0.9$}}}&	\textsf{bias}&	-0.816	&	-0.819	&	-0.023	&	-0.428	&	1.154	&	0.598	&	0.039	&	0.552	&	-1.000	&	-0.979	&	-0.039	&	-0.742	&	1.324	&	0.875	&	0.055	&	0.946	\\
&	\textsf{std}	&	0.108	&	0.108	&	0.210	&	0.184	&	0.117	&	0.279	&	0.248	&	0.250	&	0.105	&	0.105	&	0.338	&	0.200	&	0.129	&	0.340	&	0.431	&	0.273	\\
&	\textsf{rmse}	&	0.823	&	0.826	&	0.212	&	0.466	&	1.160	&	0.660	&	0.251	&	0.606	&	1.005	&	0.985	&	0.340	&	0.769	&	1.331	&	0.938	&	0.435	&	0.985	\\
&	\textsf{size}	&	1.000	&	1.000	&	0.055	&	0.645	&	1.000	&	0.540	&	0.051	&	0.599	&	1.000	&	1.000	&	0.053	&	0.958	&	1.000	&	0.733	&	0.056	&	0.639	\\
\bottomrule
\end{tabular}
\end{center}
\end{table}
\begin{table}[htbp]
\footnotesize 
\begin{center}
\caption{\small \sf DGP2}\label{table2} 
\addtolength{\tabcolsep}{-0.6em}
\begin{tabular}{clcccccccccccccccc} \toprule
&& \multicolumn{8}{c}{$\Gamma(1,1)$} & \multicolumn{8}{c}{$\Gamma(3,2)$}\\
 \cmidrule(l){3-10} \cmidrule(l){11-18} 
&& \multicolumn{4}{c}{$\beta $} &  \multicolumn{4}{c}{$\gamma $} & \multicolumn{4}{c}{$\beta $} &  \multicolumn{4}{c}{$\gamma $}\\
 \cmidrule(l){3-6} \cmidrule(l){7-10} \cmidrule(l){11-14} \cmidrule(l){15-18}
\multicolumn{2}{l}{$\alpha=0$} & \sf OLS  &   \sf GP   &  \sf npCF  & \sf 2sC &   \sf OLS  &  \sf GP   & \sf npCF  & \sf 2sC & \sf OLS  &   \sf GP   &  \sf npCF  & \sf 2sC &   \sf OLS  &  \sf GP   & \sf npCF  & \sf 2sC\\ \hline
\parbox[t]{2mm}{\multirow{4}{*}{\rotatebox[origin=c]{90}{$\rho =0.0$}}}&	\textsf{bias}&	0.000	&	0.000	&	-0.001	&	-0.001	&	0.002	&	0.004	&	0.002	&	0.004	&	0.000	&	0.000	&	-0.002	&	-0.001	&	0.002	&	0.004	&	0.006	&	0.007	\\
&	\textsf{std}	&	0.065	&	0.065	&	0.067	&	0.065	&	0.064	&	0.168	&	0.154	&	0.150	&	0.065	&	0.065	&	0.067	&	0.066	&	0.074	&	0.309	&	0.283	&	0.277	\\
&	\textsf{rmse}	&	0.065	&	0.065	&	0.067	&	0.065	&	0.064	&	0.168	&	0.154	&	0.150	&	0.065	&	0.065	&	0.067	&	0.066	&	0.074	&	0.309	&	0.283	&	0.278	\\
&	\textsf{size}	&	0.044	&	0.059	&	0.054	&	0.047	&	0.049	&	0.036	&	0.033	&	0.050	&	0.044	&	0.057	&	0.047	&	0.045	&	0.043	&	0.04	&	0.039	&	0.048	\\
\cmidrule(l){3-10}\cmidrule(l){11-18}
\parbox[t]{2mm}{\multirow{4}{*}{\rotatebox[origin=c]{90}{$\rho =0.5$}}}&	\textsf{bias}&	-0.001	&	0.000	&	0.000	&	0.000	&	0.457	&	-0.016	&	0.020	&	0.014	&	-0.001	&	0.000	&	-0.001	&	-0.001	&	0.559	&	0.079	&	0.031	&	0.036	\\
&	\textsf{std}	&	0.057	&	0.056	&	0.080	&	0.062	&	0.065	&	0.153	&	0.138	&	0.136	&	0.056	&	0.056	&	0.068	&	0.062	&	0.069	&	0.282	&	0.255	&	0.252	\\
&	\textsf{rmse}	&	0.057	&	0.056	&	0.080	&	0.062	&	0.461	&	0.154	&	0.14	&	0.137	&	0.056	&	0.056	&	0.068	&	0.062	&	0.564	&	0.293	&	0.257	&	0.254	\\
&	\textsf{size}	&	0.027	&	0.055	&	0.054	&	0.047	&	1.000	&	0.033	&	0.061	&	0.037	&	0.021	&	0.055	&	0.040	&	0.033	&	1.000	&	0.058	&	0.057	&	0.055	\\
\vspace*{-.32cm}\\
\cmidrule(l){3-10}\cmidrule(l){11-18}
\multicolumn{2}{l}{$\alpha=0.5$} & \sf OLS  &   \sf GP   &  \sf npCF  & \sf 2sC &   \sf OLS  &  \sf GP   & \sf npCF  & \sf 2sC & \sf OLS  &   \sf GP   &  \sf npCF  & \sf 2sC &   \sf OLS  &  \sf GP   & \sf npCF  & \sf 2sC\\ \hline
\parbox[t]{2mm}{\multirow{4}{*}{\rotatebox[origin=c]{90}{$\rho=0.0$}}}&	\textsf{bias}&	0.000	&	-0.001	&	0.000	&	-0.002	&	-0.001	&	-0.002	&	0.000	&	0.002	&	-0.001	&	-0.001	&	-0.001	&	-0.003	&	-0.001	&	-0.003	&	0.001	&	0.004	\\
&	\textsf{std}	&	0.073	&	0.074	&	0.134	&	0.094	&	0.074	&	0.169	&	0.210	&	0.150	&	0.073	&	0.074	&	0.179	&	0.120	&	0.085	&	0.309	&	0.370	&	0.248	\\
&	\textsf{rmse}	&	0.073	&	0.074	&	0.134	&	0.094	&	0.074	&	0.169	&	0.210	&	0.150	&	0.073	&	0.074	&	0.179	&	0.120	&	0.085	&	0.309	&	0.370	&	0.249	\\
&	\textsf{size}	&	0.046	&	0.051	&	0.043	&	0.047	&	0.047	&	0.033	&	0.054	&	0.048	&	0.050	&	0.051	&	0.037	&	0.048	&	0.048	&	0.041	&	0.040	&	0.048	\\
\cmidrule(l){3-10}\cmidrule(l){11-18}
\parbox[t]{2mm}{\multirow{3}{*}{\rotatebox[origin=c]{90}{$\rho =0.5$}}}&	\textsf{bias}&	-0.262	&	-0.287	&	0.056	&	-0.014	&	0.576	&	0.054	&	0.008	&	0.023	&	-0.289	&	-0.291	&	0.003	&	-0.026	&	0.714	&	0.222	&	0.048	&	0.055	\\
&	\textsf{st.err.}	&	0.064	&	0.064	&	0.127	&	0.086	&	0.071	&	0.152	&	0.194	&	0.135	&	0.063	&	0.063	&	0.166	&	0.103	&	0.074	&	0.274	&	0.346	&	0.215	\\
&	\textsf{RMSE}	&	0.269	&	0.294	&	0.139	&	0.087	&	0.580	&	0.161	&	0.195	&	0.137	&	0.295	&	0.298	&	0.166	&	0.106	&	0.718	&	0.353	&	0.349	&	0.222	\\
&	\textsf{rej.}H$_0$	&	0.977	&	0.996	&	0.063	&	0.054	&	1.000	&	0.059	&	0.042	&	0.048	&	0.990	&	0.996	&	0.041	&	0.030	&	1.000	&	0.139	&	0.052	&	0.045	\\
\bottomrule
\end{tabular}
\end{center}
\end{table}

 For all Monte Carlo experiments, we draw $n=250$ \textsf{IID} random draws $\{x_i,z_i,u_i\}_{i=1}^n$ from $(x,z,u)$ according to {\sf DGP1} and {\sf DGP2}. The proposed nonparametric control function estimator ({\sf npCF}) is compared with OLS (i.e. ignoring the endogeneity of $z$), with the estimator by \citet{guptapark:2012} ({\sf GP}) and with the {\sf 2sCOPE} estimator by \citet{yang:2022}.\footnote{We employ the {\sf R}-package `REndo' for computing {\sf GP}. All computations were parallelized and performed using CHEOPS, the DFG-funded (Funding number: INST 216/512/1FUGG) High Performance Computing (HPC) system of the Regional Computing Center at the University of Cologne (RRZK) using  iterations.} 
The main difference between {\sf 2sCOPE} and our method is that the rank transformations as in \eqref{ranktransformation} of the present paper are applied  already in the first step to both $x$ and $z$. Accordingly, in the first step the transformed endogenous regressor ($\widehat \eta$) is regressed on the corresponding transformations of the exogenous regressors. In the second step, the OLS residuals of this regression are added to the original equation. This procedure is designed to capture the dependence among the regressors as in {\sf DGP2}, but the procedure is expected to be less suitable for {\sf DGP1}.
 
In order to investigate the validity of statistical inference we consider $t$-type test statistics for the hypotheses $\beta_1 = -1$ and $\gamma=1$ respectively at a nominal significance level of 5\%. For the OLS estimator the usual $t$-statistics are used, whereas the other tests are performed using \(t\)-statistics equipped with bootstrap standard errors. Tests based on bootstrap percentile intervals did perform very similar so that the corresponding results are omitted to save space. Our experience suggests that using 99 bootstrap replications is typically enough to obtain reliable estimates for the unknown standard deviations of the coefficients. Finally, all results are based on 1,000 Monte Carlo repetitions.

 {\bf Discussion of \sf DGP1.} We first discuss the results in Table \ref{table1} for the case $e \sim \Gamma(1,1)$ if $\delta=0$ . That is, the setup with uncorrelated regressors that is assumed by \citet{guptapark:2012}. Not surprisingly, OLS performs best in case of no endogeneity with a RMSE less than the half of the other three approaches. This highlights the efficiency loss resulting from the endogeneity correction if it is in fact not necessary. A similar efficiency loss occurs when employing external instruments.\footnote{It should be noted, however, that it is easy to verify that the endogeneity correction is unnecessary by observing an insignificant $t$-statistic for the correction term $\widehat \eta_i$.} Note that in this case the asymptotic distribution of the latter IV estimator with $v$ as some external instrument is given by
\begin{align*}
\sqrt{n} (\widehat \gamma_{\textsf{IV}} -\gamma) &\stackrel{d}{\to } {\cal N}(0,V_{\textsf{OLS}}/r_{zv}^2),
\end{align*}
where $V_{\textsf{OLS}}$ is the asymptotic variance of the OLS estimator and $r_{zv}^2$ is the $R^2$ from the (first stage) regression of $z$ on $v$. Accordingly, a comparable loss of efficiency results from an IV estimator if the first stage $R^2$ corresponds to $(0.062/0.147)^2 = 0.178$.

In the presence of endogeneity (i.e. $\rho\ne 0$) the OLS estimator of the parameter $\gamma$ is highly biased. Note that for the case of uncorrelated regressors, the coefficient $\beta$ can be estimated unbiasedly even if the regressor $z$ is endogenous. Accordingly, all estimators (including OLS) for $\beta$ are unbiased in this case. On the other hand, the OLS estimator for $\gamma$ is severely biased as the corresponding regressor is endogenous. The copula based endogeneity corrections {\sf GP} and {\sf 2sCOPE} and our {\sf npCF} estimator effectively remove the bias and perform more or less similarly. Furthermore, the empirical sizes of the (bootstrap) $t$-tests are close to the nominal size. Although not reported to save space, using classical OLS standard errors resulted in case of {\sf npCF} in severe size distortions.

The case of correlated regressors with $\delta=1$ is considered in the lower panel of Table \ref{table1}. As expected, in this case the OLS estimator is biased for both coefficients $\beta$ and $\gamma$. The bias of the {\sf GP} estimator is similar to the OLS bias for $\beta$ and only slightly smaller for $\gamma$. A substantial bias reduction is obtained by applying the {\sf 2sCOPE} estimator but some bias remains in particular if $\rho=0.9$. This is due to the fact that the {\sf 2sCOPE} estimator assumes a particular nonlinear dependence between $x$ and $z$, whereas in our data generating process the dependence between $z$ and $x$ is linear. The {\sf npCF} estimator is the only estimator that efficiently removes the bias from both parameters.

Turning to the case where $e \sim \Gamma(3,2)$, we expect that the endogeneity corrections have problems to empirically identify the endogenous component of the error term. Indeed, the standard deviations of the endogeneity corrected standard deviations for the estimators of $\gamma$ increase about 50 percent. Apart from the higher standard deviations the findings are qualitatively similar to the results presented in Table \ref{table1}. It is interesting to note, however, that the magnitude of the bias of the {\sf 2sCOPE} estimator is similar to the OLS estimator suggesting that this estimator is not able to cope with the (linear) correlation between the regressors $x$ and $z$.

 {\bf Discussion of \sf DGP2.} Similar to {\sf DGP1}, the results for {\sf DGP2} confirm the shortcomings of {\sf GP} in case of correlated regressors ($\alpha \neq 0$ and $\rho=0.5$). Given that {\sf DGP2} mimics the model framework in \cite{yang:2022} that corresponds to the model assumptions of {\sf 2sCOPE}, it is not surprising that {\sf 2sCOPE} performs best in this setting.  Interestingly, our {\sf npCF} estimator effectively removes the bias even in the setting of {\sf DGP2}, whereas the {\sf 2sCOPE} estimator fails to remove the bias resulting from {\sf DGP1}. The standard deviations for {\sf DGP2} are however somewhat smaller for the {\sf 2sCOPE} estimator compared to the {\sf npCF} estimator due to the fact that the former estimator takes care of the particular nonlinear dependence in the {\sf DGP2}.  

\section{Empirical Application}\label{sec:Application}

In this section, we apply our new estimator to empirical wage data, thereby revisiting the classical economic problem of estimating the returns to schooling (see e.g. \citealp{harmon:2000}). Similarly as in \citet{chernozhukov2013inference} and \citet{rothewied:2013}, we analyze micro-level data from the US Current Population Survey in 1988. The sample size is $n=144,750$ and the model is given by
\begin{equation}\label{empiricalmodel}
\begin{split}
\textsf{ln}(\textsf{wage}_i) =  \beta_0  \ + \ & \gamma \, \textsf{educ}_i + \beta_1 \, \textsf{exper}_i + \beta_2 \, \textsf{exper}^2_i  + \beta_3 \,  \textsf{married}_i  \\
              \ + \ & \beta_4 \, \textsf{parttime}_i + \beta_5 \, \textsf{union}_i + \beta_6 \, \textsf{smsa}_i + \beta_7 \, \textsf{nonwhite}_i + u_i.   		
\end{split}
\end{equation}
Accordingly, we consider a linear regression in which the logarithm of hourly wages for individual $i$ is explained by the years of education and different control variables. The latter consist of the years of working experience (linearly and quadratically) and several binary variables (married, working in part time, member in a union, living in a standard metropolitan statistical area, being non-white). The control variables are assumed to be exogenous. The years of education is assumed to be an endogenous variable as it might be correlated with unobservable worker's characteristics in the error term $u_i$ such as ability or motivation. The key goal is to find a reliable estimate for $\gamma$.

The literature in labor economics discusses several approaches for estimating $\gamma$ and provides different empirical results for different data sets. For example, if \textsf{educ}$_i$ is instrumented with family characteristics such as the years of education of the parents, the IV estimates are often smaller than the OLS estimates (see \citealp{wooldridge:textbook}). This fits to the economic intuition that years of education might be positively correlated with ability/motivation. If institutional characteristics are used as instruments, the IV estimates are often larger (see \citealp{card:2000}). \citet{lemke:2003} try to merge these results with the conclusion that the OLS estimators might be biased upwards.

\begin{figure}
    \centering
    \subfloat[Histogram of \textit{educ}]{\includegraphics[width = .5\textwidth]{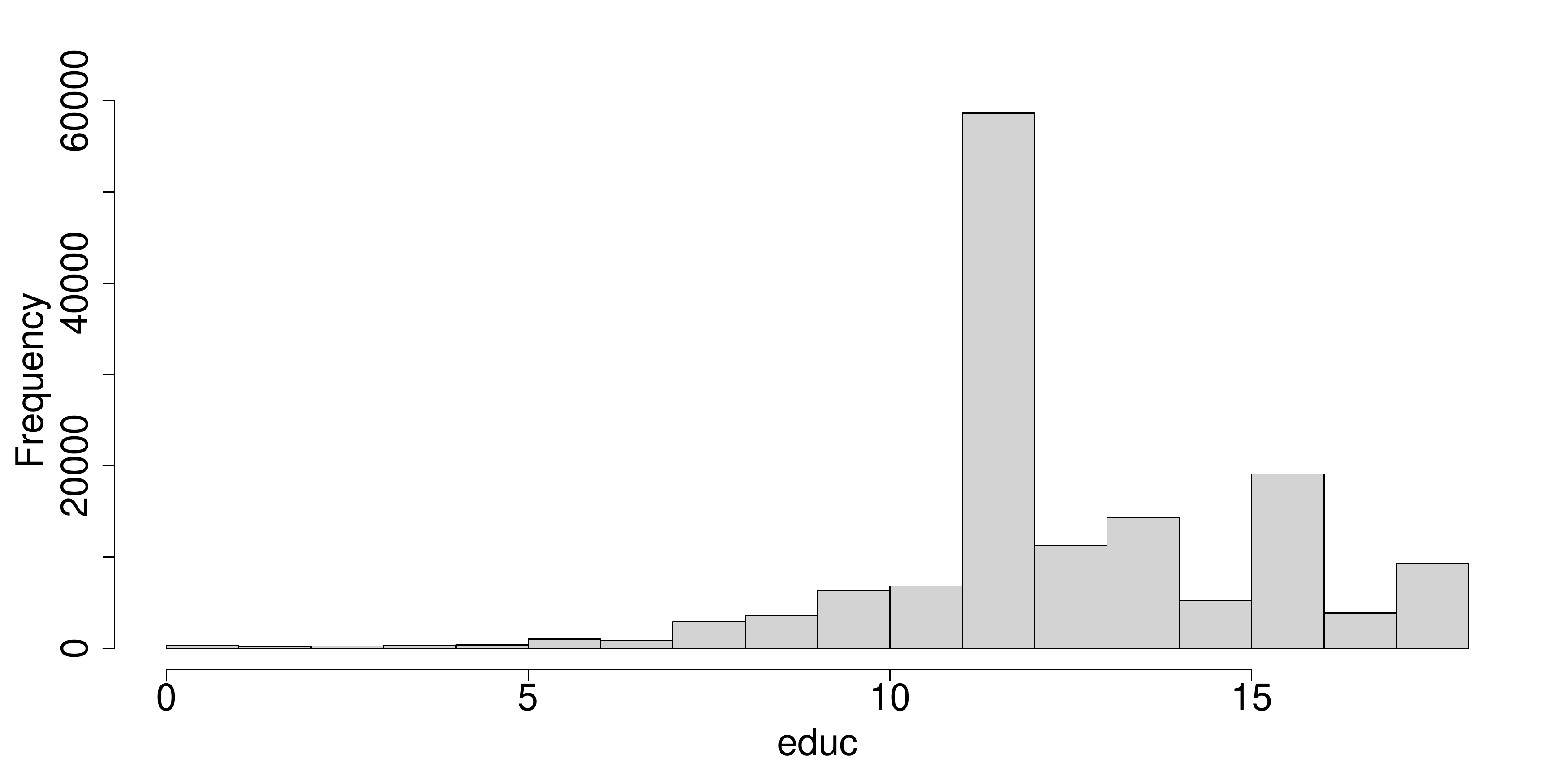}}
    \subfloat[Histogram of $\widehat{\textsf{educ}}$]{\includegraphics[width = .5\textwidth]{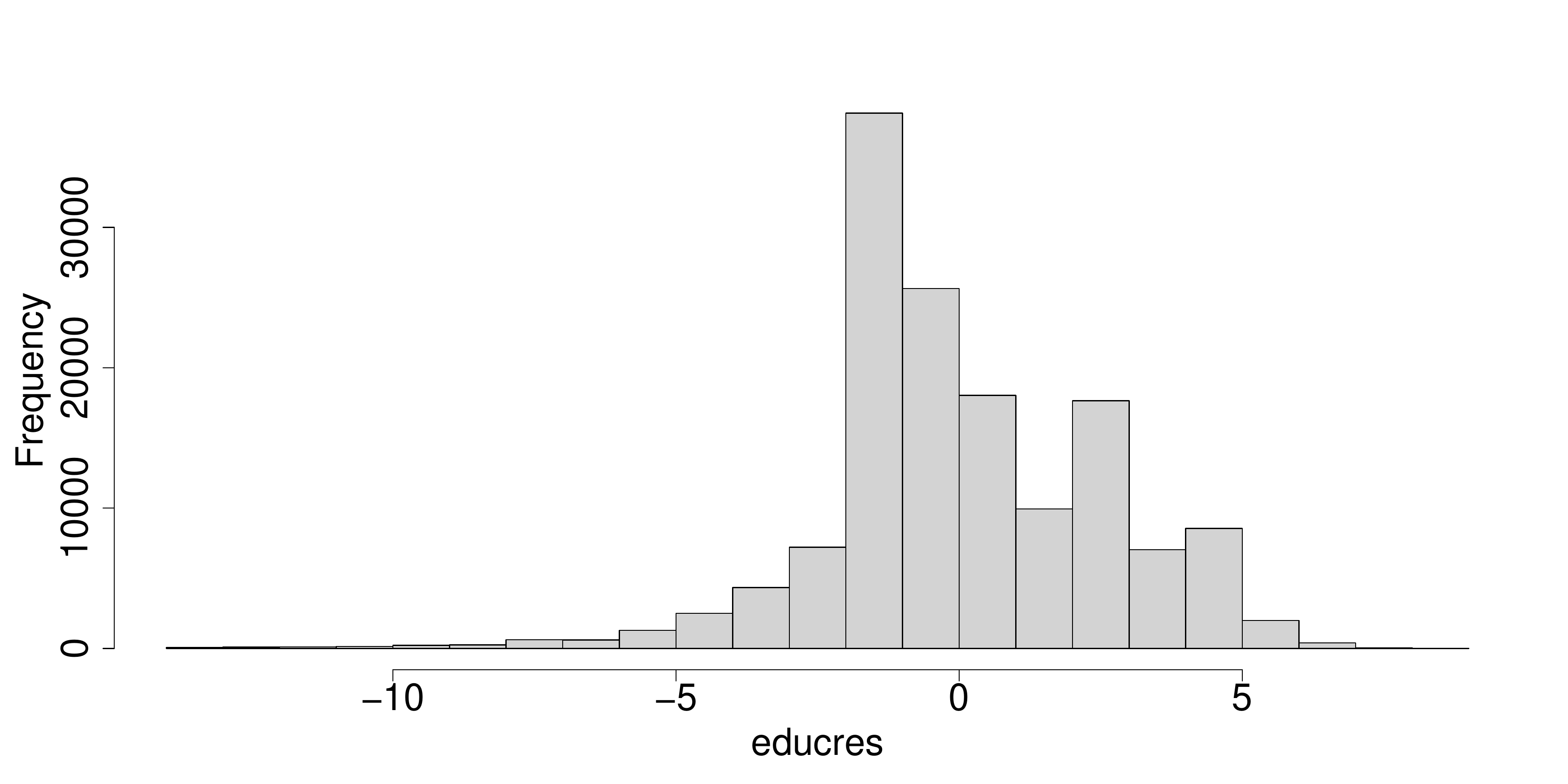}}
    \caption{Comparison of the distributions of \textsf{educ} and its $\widehat{\textsf{educ}}$}\label{figure:app}
\end{figure}

This analysis contributes to this discussion by providing results of a consistent estimator for $\gamma$ that do not depend on external instruments. The starting point is the observation that the distribution of the regressor \textsf{educ} is obviously not normal, as the histogram in Figure \ref{figure:app} (left) shows. On the one hand, the distribution is discrete and on the other hand, it is highly skewed. This is a simple argument for justifying the augmentation of the regression \eqref{empiricalmodel} by the normalizing transformation of \textsf{educ} as discussed earlier in the paper. In the first step, residuals $\hat e_i$ from an OLS regression of \textit{educ} on all exogenous control variables are obtained. Also the distribution of these residuals is highly skewed, as Figure \ref{figure:app} (right) shows. In the second step, the augmented regressor $\widehat \eta_i = \Phi^{-1}(\widehat F_{\hat e}(\widehat e_i) )$ is calculated. The distribution of $\widehat \eta_i$ is approximately normal by construction. It is the difference between the two distributions which enables identification of $\gamma$.

Note that at this point, we assume that also the latent counterpart $\eta$, which corresponds to personal characteristics such as intelligence, other skills and diligence, is normally distributed. In our point of view, this makes sense for several reasons. On the one hand, the single characteristics can be assumed as being normally distributed, e.g. intelligence: While standardized intelligence quotient (IQ) ratios are by construction normally distributed, this is less clear for the intelligence itself. However, early research in psychology about intelligence in the beginning of the 20th century (see e.g. \citealp{burt:17}), provided evidence that also intelligence is normally distributed. In later research, psychologists found some more mixed evidence, in particular the question arose whether the actual tails of the intelligence distribution might be heavier than expected under the normal distribution assumption. Nevertheless, there are also more recent studies which still support the normality assumption, see e.g. \cite{warne:13}.

Moreover, one can rely on an argument based on the central limit theorem: If several sources have an additive influence on $\eta$, the sum of these sources can be expected to be approximately normal. On the other hand, this latent variable is not a linear transformation of the years of education due to rules concerning compulsory school attendance, for example.

Table \ref{table:app} presents the results (estimates, standard errors and $t$-statistics) of the standard OLS estimation of model \eqref{empiricalmodel} and the results of the robustified estimation, where $\widehat \eta_i$ is included. In the latter case, we report both the bootstrap standard errors and $t$-statistics as well as the OLS $t$-statistics. In the latter case, the estimation error for $\widehat \eta_i$ is ignored, but the differences are small and not systematic. Due to the large sample size, the $p$-values for all coefficients are smaller than $2 \cdot 10^{-16}$. One exception is the coefficient $\rho_{\eta}$ for $\widehat \eta_i$ which is somewhat larger with $2.5 \cdot 10^{-5}$. This reflects that, on the one hand, all control variables were used to calculate $\widehat \eta_i$, but, on the other hand, $\widehat \eta_i$ still provides considerable explanatory power. The latter observation implies that it is reasonable to assume that the regressor \textsf{educ} is endogenous.

In the case of OLS, we have $\hat \gamma = 0.083$, whereas we have $\hat \gamma = 0.073$ when correcting for endogeneity. So, we observe a slight reduction of the estimated return for education which is what \citet{lemke:2003} also suggest. The other coefficients remain very similar, which could have been expected as the residuals $\hat e_i$ are orthogonal to the exogenous regressors and the adjusted $R^2$ of the first stage regression is rather low ($0.1132$). For all coefficients, the $t$-statistics in absolute values are lower and the standard errors are higher in the augmented regression compared to OLS. This is in particular the case for the intercept and $\gamma$ and can be explained with the high correlation ($0.91$) of \textsf{educ} and $\widehat \eta_i$. A similar phenomenon is observed in other studies of endogeneity-corrected estimators such as \citet{guptapark:2012}. With 2sCOPE, the estimate for $\gamma$ is less than half as large compared to npCF and the estimate for $\rho$ is more than six times larger. A direct comparison is difficult, however, as $\rho$ has a different interpretation for both estimation approaches.

\begin{table}\footnotesize
\begin{center}
\caption{\small Estimation results}  \label{table:app}
\vspace*{.1cm}
\addtolength{\tabcolsep}{-0.4em}
\begin{tabular}{rcccccc} \toprule \\[-3ex]
 & \multicolumn{2}{c}{\sf OLS} & \multicolumn{2}{c}{ \sf npCF} & \multicolumn{2}{c}{ \sf 2sCOPE} \\ 
 \cmidrule(l){2-3} \cmidrule(l){4-5} \cmidrule(l){6-7}
            & \textsf{estimate} & \textsf{t stat}  &  \textsf{estimate} & \textsf{t stat} &     \textsf{estimate} & \textsf{t stat} \\ \hline
$\gamma$ & 0.083 & 81.44 & 0.072 & 25.02 & 0.032 & 11.14  \\[-1ex]
         & {\scriptsize ($<$0.001)} &  & {\scriptsize (0.003)}  &  & {\scriptsize (0.003)} &   \\
   \cmidrule(l){2-7}
  $\beta_0$ & 0.559 & 180.31 & 0.689 & 18.50 & 1.236 & 31.37 \\[-1ex]
            & {\scriptsize (0.007)} &  & {\scriptsize (0.037)}  &  & {\scriptsize (0.039)} &   \\
  $\beta_1$ & 0.027 &  78.71 & 0.028 & 85.16 & 0.024 & 53.50 \\[-1ex]
            & {\scriptsize ($<$ 0.001)} &  & {\scriptsize ($<$ 0.001)}  &  & {\scriptsize ($<$ 0.001)} &   \\
  $\beta_2$ & $>$ -0.001 &  -54.14 & $>$ -0.001 & -49.05 & $>$ -0.001 & -47.92 \\[-1ex]
            & {\scriptsize ($<$ 0.001)} &  & {\scriptsize ($<$ 0.001)}  &  & {\scriptsize ($<$ 0.001)} &   \\
  $\beta_3$ & 0.233 &   74.91 & 0.234 & 86.31 & 0.240 & 90.14 \\[-1ex]
            & {\scriptsize (0.003)} &  & {\scriptsize (0.003)}  &  & {\scriptsize (0.003)} &   \\
  $\beta_4$ & 0.119  &   46.09 & 0.124 & 39.44 & 0.154 & 43.69 \\[-1ex]
            & {\scriptsize (0.003)} &  & {\scriptsize (0.003)}  &  & {\scriptsize (0.004)} &   \\
  $\beta_5$ & -0.333  &   -107.23 & -0.341 & -99.98 & -0.379 & -102.59 \\[-1ex]
            & {\scriptsize (0.003)} &  & {\scriptsize (0.003)}  &  & {\scriptsize (0.004)} &   \\
  $\beta_6$ & -0.114   &   -32.79 & -0.118 & -33.80 & -0.131 & -36.61 \\[-1ex]
            & {\scriptsize (0.003)} &  & {\scriptsize (0.004)}  &  & {\scriptsize (0.004)} &   \\
  $\beta_7$ & 0.181   &   69.72 & 0.188 & 64.83 & 0.208 & 73.36 \\[-1ex]
            & {\scriptsize (0.003)} &  & {\scriptsize (0.003)}  &  & {\scriptsize (0.003)} &   \\
\cmidrule(l){2-7}
  $\rho$ &   &   & 0.026 & 3.78 & 0.143 & 18.34 \\[-1ex]
            &  &  & {\scriptsize (0.007)}  &  & {\scriptsize (0.008)} & \\
\cmidrule(l){2-7} 
$R^2$ & \multicolumn{2}{c}{0.415 }  &  \multicolumn{2}{c}{0.415 }  & \multicolumn{2}{c}{0.417 } \\ 
\bottomrule 
\multicolumn{7}{p{8cm}}{\scriptsize {\textsf{Note}:} Standard errors, reported below point estimates, are computed based on 100 bootstrap replications in case of {\sf np-CF} and {\sf 2sCOPE}.}
\end{tabular}
\end{center}
\end{table}

\section{Concluding Remarks}
This paper proposes a method for solving the endogeneity problem without external instruments or specifying a copula. The endogeneity correction is obtained by simply augmenting the regression with a conformably transformed regressor. We study the asymptotic properties and analyse the small sample properties of the resulting estimator. Our results suggest that the proposed estimation framework may provide a useful tool for estimating regression models with endogenous regressors if no suitable instruments are available. Even if instrumental variables are at hand, additional nonlinear instruments can be constructed for improve the efficiency of the IV estimator. Furthermore, the nonparametric control function may be employed for testing the exogeneity of the regressors in empirical situations where valid instruments are missing.

A drawback of our approach is, however, that the distribution of the endogenous regressor needs to be  different from that of the latent variable. If the distributions are close, then the estimator suffers from multicollinearity among the endogenous regressor and the correction term resulting in potentially large standard errors. It would be interesting to develop more diagnostic tests that allow for checking how appropriate the new method is for a particular data set. An alternative strategy would relax the normality condition of \(f(e)\) and specify instead a \textit{class} of permissible distributions indexed by some finite dimensional parameter \(\tau\), say. Using nonlinear estimation techniques, one could then estimate \(\tau\) alongside the remaining model parameters \(\theta\).

\section{ACKNOWLEDGEMENTS}

We are grateful to the co-editor, Petra Todd, and two referees, as well as to Julien Bergeot and Rouven E. Haschka for helpful comments.
\bibliography{bibl} 




    \section*{Appendix A: Auxiliary Lemmata}
    \renewcommand{\theequation}{A.\arabic{equation}}
    \renewcommand{\thesection}{A}
    \setcounter{equation}{0}

    To begin with, we state two auxiliary results. The first lemma is due to \cite{zhao:2020} and concerns the difference between the residual rank \(\hhFe\) and its `oracle' counterpart \(\hFe\) based on the ecdf $\hat{F}_e(x) \coloneqq (n+1)^{-1}\sum_{i = 1}^n1\{e_i \leq x\}.$ For future reference, let us also introduce the $n \times 1$ vector of the `oracle' quantile transforms $\tilde\eta \coloneqq (\tilde\eta_1,\dots,\tilde\eta_n)'$, with $\tilde\eta_i \coloneqq \Phi^{-1}(\hFe)$.
\renewcommand{\thelemma}{A}
\begin{lemma}\label{lA}  For any \(i = 1,\dots,n\),
\[\normalfont
\hhFe-\hFe = - \sqrt{n}(\hat \delta-\delta)'(x_i-\mu_x) f_e(e_i) + R_i,\quad \mmax\limits_{1 \leq i \leq n} |R_i| = o_p(n^{-1/2}).
\]
\end{lemma}

\renewcommand{\thelemma}{B}
\begin{lemma}\label{lB}  Let \(F\) be a continuous cdf with \(\int_0^1(F^{-1}_i(u))^2du < \infty\), then
\begin{align*}\normalfont
\int_0^1\int_0^1 \frac{F^{-1}(u)}{\phi(\Phi^{-1}(u))}\frac{\Phi^{-1}(v)}{\phi(\Phi^{-1}(v))}(\textsf{min}(u,v)-uv) dudv =  \frac{1}{2} \int_0^1F^{-1}(u)\Phi^{-1}(u)du.
\end{align*}
\end{lemma}

 \noindent\textbf{Proof of Lemma \ref{lA}.} This is \citet[Prop. 3.2]{zhao:2020}. \hfill$\square$\\
 
\noindent\textbf{Proof of Lemma \ref{lB}.} Let \(g(u) \coloneqq \phi(\Phi^{-1}(u))\) and consider
\begin{align*}
\int_0^1\int_0^1 \frac{F^{-1}(u)}{\phi(\Phi^{-1}(u))}&\frac{\Phi^{-1}(v)}{\phi(\Phi^{-1}(v))}(\textsf{min}(u,v)-uv) dudv\\
 =\,& \int_0^1F^{-1}(u)  \left(\int_0^1 \frac{\Phi^{-1}(v)}{\phi(\Phi^{-1}(v)) \phi(\Phi^{-1}(u))}(\textsf{min}(u,v)-uv) dv\right)du.
\end{align*}
Numerical integration reveals for any \(u \in (0,1)\)
\[
\int_0^1 \frac{\Phi^{-1}(v)(\textsf{min}(u,v)-uv)}{\phi(\Phi^{-1}(v))} dv = \frac{\Phi^{-1}(u)}{2}\int_0^1 \frac{\textsf{min}(u,v)-uv}{\phi(\Phi^{-1}(v))} dv.
\]
Combining the above with the fact that
\[
\int_0^1\int_0^1\frac{\textsf{min}(u,v)-uv}{\phi(\Phi^{-1}(v)) \phi(\Phi^{-1}(u))}dudv = 1
\]
finishes the proof. \hfill$\square$\\

    \section*{Appendix B: Proof of Main Results}
    \renewcommand{\theequation}{B.\arabic{equation}}
    \renewcommand{\thesection}{B}
    \setcounter{equation}{0}
    \subsection{Proof of Theorem 2.1}
To begin with, note that monotonicity of $f(\cdot)$ and normality of $f(e)$ implies $F_e(a) = \PP(e \leq a) = \PP(f(e) \leq f(a)) = \Phi(f(a)) \Leftrightarrow f(a) = \Phi^{-1}(F_e(a))$, where $F_e(\cdot)$ is the c.d.f. of $e$. First, suppose $(\beta',\gamma,\rho)'$ are not identified such that there exists some linear combination $\lambda_1'x + \lambda_2 z + \lambda_3  f(e) = 0$, with $\lambda_j \neq 0$ for at least one index $j \in \{1,2,3\}$. Since
\begin{equation}\label{proof21}
\lambda_1'x + \lambda_2 z + \lambda_3 f(e) = 0 \Rightarrow \lambda_1'\E[x] + \lambda_2 \delta'\E[x] + \lambda_2 e + \lambda_3 f(e)  = 0 
\end{equation}
Since $\Var[e]>0$, the above can only hold ($a.s.$) if $\lambda_j \neq 0$, $j \in \{2,3\}$. Next, upon applying the standard normal c.d.f. to the preceding display, we get
\[
\Phi\left(-\frac{(\lambda_1'\E[x] + \lambda_2 \delta'\E[x])}{\lambda_3} -\frac{\lambda_2}{\lambda_3} e\right) = F_e(e). 
\]
Because $F_e(e) \sim {\sf UNIF}[0,1]$, this can only be true if $e$ is normal. However, if we assume that $e$ is normal, then \eqref{proof21} is solved for  $\lambda_2 = -\lambda_3$ and $\lambda_1 = \lambda_3\delta$. \hfill$\square$\\

\subsection{Proof of Proposition 3.1}
First, we show that \((W'W/n)\sqrt{n}(\hat\theta-\theta) = \Lambda_n+o_p(1)\), where
\begin{align}\label{eq:asydecom}
\Lambda_n = \begin{bmatrix} \Lambda_{1,n}\\\delta'\Lambda_{1,n}+\lambda_{1,n}\\\lambda_{2,n}\end{bmatrix},
\end{align}
with \(\Lambda_{1,n} \coloneqq X'(\eps + \rho c_1 e)/\sqrt{n}\), \(\lambda_{1,n} \coloneqq e'(\eps+\rho(\eta-\tilde\eta))/\sqrt{n}\), \(\lambda_{2,n} \coloneqq \eta'(\eps+\rho(\eta-\tilde\eta))/\sqrt{n}\).
Without loss of generality, we drop the intercept and assume that all variables are centred. In view of Eq. \eqref{eq:thetadecom}, we note that
\(
W'\eps/\sqrt{n} = (\eps'X, \delta'X'\eps + e'\eps, \hat\eta'\eps)'/\sqrt{n}.
\)
Next, we show that
\[
W'(\eta-\hat\eta)/\sqrt{n} = (c_1e'X, c_1\delta'X'e+e'(\eta-\tilde\eta), \eta'(\eta-\tilde\eta))'/\sqrt{n} + o_p(1),
\]
which follows upon consecutively investigating the three terms \(A_n \coloneqq n^{-1/2}X'(\eta-\hat\eta)\), \(B_n \coloneqq n^{-1/2}Z'(\eta-\hat\eta)\), and \(C_n \coloneqq n^{-1/2}\hat\eta'(\eta-\hat\eta)\), referred to respectively as `Step A', `Step B', and `Step C'.  

\noindent {\bf Step A.} Let \(A_n \coloneqq A_{1,n}-A_{2,n}\), with \(A_{1,n} \coloneqq X'(\eta-\tilde\eta)/\sqrt{n}\) and \(A_{2,n} \coloneqq X'(\hat\eta-\tilde\eta)/\sqrt{n}\).
 We first show that \(A_{1,n} = o_p(1)\). To do so, note that there are \(i.i.d.\) variates \(U_i \sim \textsf{UNIF}[0,1]\), \(i \in \{1,\dots,n\}\), such that \(e_i = F_e^{-1}(U_i)\). Let \(F_n\) denote the empirical cdf of \(U_1,\dots,U_n\).  There exists an ordering of the indices \(\{1,\dots,n\}\) such that
\begin{align*}
A_{1,n} =   \frac{1}{\sqrt{n}}\sum_{i=1}^n x_i[\Phi^{-1}(F_n^{-1}(i/n))-\Phi^{-1}(i/\tn)] =  \frac{1}{\sqrt{n}}\sum_{i=1}^n x_i\frac{F_n^{-1}(i/\tn)-i/\tn}{\phi(\Phi^{-1}(i/\tn))} + o_p(1),
\end{align*}
where $\tn \coloneqq n+1$. The second equality uses a Taylor expansion around \(i/\tn\) and the order of the remainder term is due to arguments used in the proof of \citet[Thm. 3.4.]{zhao:2020}. Next, the weak convergence of the uniform quantile process \(\sqrt{n}(F_n^{-1}(u)-u),\) \(u \in (0,1)\), reveals that
\begin{align}\label{eq:weakquant}
\frac{1}{\sqrt{n}}\sum_{i=1}^n \frac{F_n^{-1}(i/\tn)-i/\tn}{\phi(\Phi^{-1}(i/\tn))} \stackrel{d}{\rightarrow} \int_0^1 \frac{B(u)}{\phi(\Phi^{-1}(u))}du \eqqcolon H,
\end{align}
for a standard Brownian bridge $B(\cdot)$; see e.g. \cite{koenkerxiao:2002}. Results collected in \citet{webelwied:2016} yield that $H$ is normally distributed with expectation zero and variance
 \[
\Var[H] = \int_0^1\int_0^1  h(s) h(u) (\textsf{min}(s,u)-su) ds du = 1, \quad h(u) \coloneqq 1/\phi(\Phi^{-1}(u)),
\]
using Lemma \ref{lB}. As \(x_i\) is mean-zero and independent of \(U_i\), it follows that
\[
\sum_{i=1}^n \frac{x_i}{\sqrt{n}}\frac{\sqrt{n}(F_n^{-1}(i/\tn)-i/\tn)}{\phi(\Phi^{-1}(i/\tn)} = O_p(1),
\]
so that \(A_{1,n} = o_p(1)\). Turning to \(A_{2,n}\), we get from a first-order Taylor expansion in conjunction with Lemma \ref{lA} and arguments used in the proof of \citet[Thm. 3.4.]{zhao:2020}
\begin{align}
A_{2,n} = \,& -\frac{1}{n}\sum_{i=1}^n \frac{f_e(F_e^{-1}(\hFe))}{\phi(\Phi^{-1}(\hFe))}x_ix_i'\sqrt{n}(\hat\delta-\delta)  + o_p(1) = -c_1X'e/\sqrt{n}+ o_p(1),
\end{align}
where the second equality follows from the LLN, \(\sqrt{n}(\hat\delta-\delta) = \Sigma_x^{-1}X'e/\sqrt{n}+o_p(1)\), and the independence between $x$ and $e$.

\noindent {\bf Step B.} Next, let \(B_n = B_{1,n}-B_{2,n}\),  with \(B_{1,n} \coloneqq z'(\eta-\tilde\eta)/\sqrt{n}\) and \(B_{2,n} \coloneqq z'(\hat\eta-\tilde\eta)/\sqrt{n}\) for \(z \coloneqq (z_1,\dots,z_n)'\) and recall that \(z_i = \delta'x_i + e_i\).
Since \(A_{1,n} = o_p(1)\), it follows directly that
\(
B_{1,n} =   n^{-1/2} e'(\eta-\tilde{\eta}) + \delta'A_{1,n}=  n^{-1/2} e'(\eta-\tilde{\eta}) + o_p(1).
\)
Similarly, using the same arguments used to examine \(A_{2,n}\), we get
\(
B_{2,n} = \delta'A_{2,n}+ o_p(1).
\)

\noindent {\bf Step C.} Finally, consider \(C_n = C_{1,n}-C_{2,n}\),  with \(C_{1,n} \coloneqq \hat\eta'(\eta-\tilde\eta)/\sqrt{n}\) and \(C_{2,n} \coloneqq \hat\eta'(\hat\eta-\tilde\eta)/\sqrt{n}\). Thus,
\begin{align}
C_{1,n} = \,& \eta'(\eta-\tilde\eta)/\sqrt{n}+(\hat\eta-\tilde\eta)'(\eta-\tilde\eta)/\sqrt{n}-(\eta-\tilde\eta)'(\eta-\tilde\eta)/\sqrt{n} \nonumber \\
    = \,& \eta'(\eta-\tilde\eta)/\sqrt{n}+o_p(1),
\end{align}
using repeatedly Cauchy-Schwarz's inequality in conjunction with \((\hat\eta-\tilde\eta)'(\hat\eta-\tilde\eta)/\sqrt{n} = o_p(1)\); see \citet[Lemma A.6]{zhao:2020} and, by Eq. \eqref{eq:weakquant}, \((\eta-\tilde\eta)'(\eta-\tilde\eta)/\sqrt{n} = o_p(1)\). Similarly, it follows that \(C_{2,n} = o_p(1)\). Finally, because $\E[\eps | \eta] = 0$, one gets from the above $\hat\eta'\eps/\sqrt{n} = \eta'\eps/\sqrt{n} + o_p(1)$. This verifies Eq. \eqref{eq:asydecom}.

Next, note that
\begin{align}
\frac{e'(\eta-\tilde\eta)}{\sqrt{n}} = \,& \frac{1}{\sqrt{n}}\sum_{i=1}^n F_e^{-1}(F_n^{-1}(i/\tn))\frac{F_n^{-1}(i/\tn)-i/\tn}{\phi(\Phi^{-1}(i/\tn))}+o_p(1) \\
\,& \stackrel{d}{\rightarrow}   \int_0^1h_1(u)B(u)du \eqqcolon H_1, \nonumber
\end{align}
where \(h_1(u) \coloneqq \frac{F_e^{-1}(u)}{\phi(\Phi^{-1}(u))}\). The variate $H_1$ is normally distributed with expectation zero and variance
\[
\Var[H_1] = \int_0^1\int_0^1  h_1(s) h_1(u) (\textsf{min}(s,u)-su) ds du \eqqcolon c_3.
\]
Note that, by Cauchy-Schwarz's inequality, \(c_3 \leq C\sigma_e^2 < \infty\) for some \(C \in (0,\infty)\). Moreover, by similar arguments
\begin{align}
\frac{\eta'(\eta-\tilde\eta)}{\sqrt{n}} =  \frac{1}{\sqrt{n}}\sum_{i=1}^n \Phi^{-1}(F_n^{-1}(i/\tn))\frac{F_n^{-1}(i/\tn)-i/\tn}{\phi(\Phi^{-1}(i/\tn))}+o_p(1) \stackrel{d}{\rightarrow} \int_0^1h_2(u)B(u)du \eqqcolon H_2, \nonumber
\end{align}
where \(h_2 \coloneqq \frac{F_e^{-1}(u)}{\phi(\Phi^{-1}(u))}\), while \(H_2\) is a mean-zero Gaussian variate with
\[
\Var[H_2] = \int_0^1\int_0^1  h_2(s) h_2(u) (\textsf{min}(s,u)-su) ds du = 1/2,
\]
using Lemma \ref{lB} with \(F = \Phi\). Recalling that \(\eta = \Phi^{-1}(F_e(e))\), we observe that \(\{e'(\eta-\tilde\eta)/\sqrt{n},\eta'(\eta-\tilde\eta)/\sqrt{n}\}\) converge jointly in distribution to a mean-zero Gaussian random vector with covariance \(\Cov[H_i,H_j] = c_31\{i = j = 1\}+1\{i=j=2\}/2+1\{i \neq j\}c_2/2\), where the case \(i \neq j\) follows from Lemma \ref{lB} with \(F = F_e\).
Recall \(\lambda_{1,n} \coloneqq e'(\eps+\rho(\eta-\tilde\eta))/\sqrt{n}\), \(\lambda_{2,n} \coloneqq \eta'(\eps+\rho(\eta-\tilde\eta))/\sqrt{n}\). Because $e$ and $\eps$ are mean-zero and uncorrelated, we obtain the limiting covariance
\begin{align*}
\lim\limits_{n \rightarrow \infty}\Cov[\lambda_{i,n},\lambda_{j,n}] = \,&  (\E[e^2\eps^2]+\rho^2c_3)1\{i=j=1\}\\
\,& + (\E[\eta^2\eps^2]+\rho^2/2)1\{i=j=2\}+(\E[e\eta\eps^2]+\rho^2/2)1\{i \neq j\},
\end{align*}
where $\E[e^2\eps^2]$, $\E[\eta^2\eps^2]$, and $\E[e\eta\eps^2]$ are positive and, by Assumption \ref{ass:Ae}, finite constants. Moreover, by the Lindeberg-Lévy CLT and the Cramér-Wold device, we obtain
$n^{-1/2}e'\eps \stackrel{d}{\rightarrow} \mathcal{N}(0, \E[e^2\eps^2])$, $n^{-1/2}X'\eps \stackrel{d}{\rightarrow} \mathcal{N}(0_k, \E[xx'\eps^2])$, $n^{-1/2 }X'e \stackrel{d}{\rightarrow} \mathcal{N}(0_k,\Sigma_x\sigma_e^2)$. Joint convergence follows from the previous results upon noting that the right-hand side of Eq. \eqref{eq:asydecom} can be expressed as a continuous transformation of the empirical process
\[
\hat{\mathbb{F}}_{x\eps e}(t_1,\dots,t_k,u,v)\coloneqq\sqrt{n}(\hat{F}_{x\eps e}(t_1,\dots,t_k,u,v)-F_{x\eps e}(t_1,\dots,t_k,u,v)),
\]
where $F_{x\eps e}(\cdot)$ is the joint $c.d.f$ of $(x',\eps,e)'$ and
\[
\hat{F}_{x\eps e} \coloneqq \frac{1}{n+1}\sum_{i = 1}^n1\{x_{i,1}\leq t_1,\dots,x_{i,k} \leq t_k,\eps_i \leq u, e_i \leq  v\}, \quad t_j,u,v\in \mathbb{R},
\]
while we recall that \((x_i,\eps_i,e_i)\) is an \(i.i.d.\) sequence by Assumption \ref{ass:A} (note that $e = z-\delta'x$). Since the class of distribution functions on $\mathbb{R}^d$, $d \geq 1$, is Donsker, \(\hat{\mathbb{F}}(\cdot)\) converges weakly to a tight Gaussian process in \(\ell^\infty(\mathbb{R}^{k+2})\). 
The claim follows, because, from the arguments used above, one deduces readily that \(W'W/n \stackrel{p}{\rightarrow} M\). \hfill$\square$\\

\subsection{Proof of Corollary 3.1} Let \(\Lambda_n^*\) be defined like Eq. \eqref{eq:asydecom} based on the bootstrap sample. Following the proof of Proposition \ref{prop:PA}, we can show that \(\sqrt{n}(\hat\theta^*-\hat\theta) = M^{-1}(\Lambda_n^*-\Lambda_n)+o_{\mathbb{P}_{\mathcal X}}(1).\) The claim of part ($i$) follows from the CLT for the bootstrap; see \citet[Chapter 19]{van2000asymptotic}. Part ($ii$) is a direct consequence of \citet[Lemma 1]{cheng:2015}. \hfill$\square$\\

\subsection{Proof of Corollary 3.2}
Under the null hypothesis the $t$-statistic can be written as
\begin{align*}
t  \coloneqq \frac{ \widehat \eta'\varepsilon - \widehat \eta'V(V'V)^{-1}V'\varepsilon }{ \widehat \sigma_\varepsilon \sqrt{ \widehat \eta'\widehat \eta - \widehat \eta'V(V'V)^{-1} V'\widehat \eta }}, \quad V \coloneqq (v_1',\dots,v_n')', \quad v_i \coloneqq (x_i',z_i)',
\end{align*}
where  $\widehat \sigma_\varepsilon^2$ is the usual estimator of the residual variance. Using the results from the proof of Proposition \ref{prop:PA} it is not difficult to show that $\widehat \sigma_\varepsilon^2\stackrel{p}{\to } \sigma_\varepsilon^2$, $n^{-1} \widehat \eta'\widehat \eta = n^{-1} \eta'\eta + o_p(1) \stackrel{p}{\to} 1$, and
\begin{align*}
\frac{1}{n} \widehat \eta'V(V'V)^{-1} V'\widehat \eta &= \frac{1}{n} \eta'V(V'V)^{-1} V'\eta + o_p(1).
\end{align*}
For the numerator of the $t$-statistic we use that \(\hat\eta'\varepsilon/\sqrt{n} \stackrel{d}{\rightarrow} \mathcal{N}(0,\sigma_\eps^2)\) and \(\hat\eta'V/\sqrt{n} = \eta'V/\sqrt{n} + o_p(1)\).
Collecting these results allows for representing the $t$-statistic as
\begin{align*}
t = \frac{ \eta'(I_n - V(V'V)^{-1}V')\varepsilon }{ \sigma_\varepsilon \sqrt{ \eta'(I_n - V(V'V)^{-1} V')\eta }} +o_p(1).
\end{align*}
Accordingly, under the null hypothesis $\rho=0$ the estimation error $\hat\eta - \eta$ does not enter the asymptotic distribution and the test statistic is asymptotically equivalent to the test statistic when $\eta$ is known. From standard results for the linear regression it immediately follows that the test statistic has the usual standard normal limiting distribution. \hfill$\square$\\

\end{document}